\documentclass{PoS}
\usepackage{bbold}
\usepackage{amsmath,amsfonts,epsfig,mathrsfs,todonotes,yfonts}
\usepackage{stmaryrd}
\usepackage[bbgreekl]{mathbbol}

\newcommand{\ber}{\begin{eqnarray}}
\newcommand{\eer}{\end{eqnarray}}
\newcommand{\vp}{\varphi}
\newcommand{\nn}{\nonumber}
\newcommand{\na}{\nabla}
\newcommand{\half}{{\textstyle{\frac12}}}
\newcommand{\ihalf}{{\textstyle{\frac i 2}}}
\newcommand{\re}[1] {(\ref{#1})}
\newcommand{\Lie }{{\cal L}}

\def\+{{+\!\!\!+}}

\title{Strong K\"ahler with Torsion as Generalised Geometry}

\ShortTitle{SKT}

\author{Chris Hull\\  The Blackett Laboratory, Imperial College London\\
Prince Consort Road London SW7 @AZ, U.K.\\      E-mail: \email{c.hull@imperial.ac.uk}}

\author{\speaker{Ulf Lindstr\"om}\thanks{Preprint: Uppsala UUITP-11/19, Imperial-TP-2019-CH-03.}\\
        Department of  Physics and Astronomy,\\ Division of Theoretical Physics,
Uppsala University,\\ Box 516, SE-751 20 Uppsala, Sweden,\\ and \\The Blackett Laboratory, Imperial College London\\
Prince Consort Road London SW7 @AZ, U.K.\\
        E-mail: \email{ulf.lindstrom@physics.uu.se}}

\abstract{Strong K\"ahler with Torsion is the target space geometry of $(2,1)$ and $(2,0)$ supersymmetric nonlinear sigma models. We discuss how it can be represented in terms of Generalised Complex Geometry in analogy to the Gualtieri map 
from the geometry of  $(2,2)$ supersymmetric nonlinear sigma models
to Generalised K\"ahler Geometry. }

\FullConference{Corfu Summer Institute 2018 ``School and Workshops on Elementary Particle Physics and Gravity''\
		(CORFU2018)\\
		31 August - 28 September, 2018\\
		Corfu, Greece}

\begin{document}

\section{Introduction}
\label{intro}

The target space geometry of $(2,2)$ supersymmetric nonlinear sigma models was shown a long time ago to be bi-hermitean, a generalisation of K\" ahler geomtery with torsion \cite{Gates:1984nk}. More recently this geometry was reformulated as Generalised K\"ahler Geometry  (GKG) \cite{Gualtieri:2003dx}, a class of Generalised Complex Geometry \cite{Hitchin:2004ut}. The relation between bi-hermitean geometry and Generalised K\"ahler Geometry is encoded in the Gualtieri map that relates the metric and two complex structures of the  bi-hermitan geometry to the two commuting Generalised Complex structures of GKG.

In this presentation we will discuss the extension of this to the target space geometry of $(2,0)$ supersymmetric nonlinear sigma models formulated in \cite{Hull:1985jv} and the geometry of 
$(2,1)$ models formulated in  \cite{Hull:1985zy}. The geometry of general  $(p,q)$ models   was given in \cite{Hull:1986hn} and further discussed in \cite{Howe:1996kj},
and recent progress includes \cite{Hull:2016khc} and \cite{Hull:2017hfa}.  The Strong K\"ahler with Torsion (SKT) geometry of the 
$(2,0)$ and $(2,1)$
 target spaces has a complex structure that is covariantly constant with respect to a  connection with  a torsion given by a closed 3-form
 and a hermitian metric.  We shall map these onto an (integrable) ``Half Generalised Complex Structure'' \cite{Hull:2018jkr}.

The content of this lecture has  been extended and formalised in a paper \cite{Hull:2018jkr}  to which the interested reader is referred for more details, including the   generalisation to $(p,q)$ geometries.

\section{Definitions}

We shall consider GCG as defined on the generalised target space 
\ber
\mathbb {T}:=T{\cal M}\oplus T^*{\cal M}
\eer
The natural pairing ${\cal{P}}$ defines a product $<,>$ defined for\footnote{See below for the matrix. notation} $\vp_1, \vp_2 \in \mathbb {T}$ 
\ber
<\vp_1,\vp_2> = \vp_1^t{\cal{P}}\vp_2~.
\eer
Writing 
\ber\label{element}
\vp=\left(\begin{array}{c}v_i\\ \xi_i\end{array}~,
\right)
\eer
with $v_i \in T$ and $x_i,\in T*$, it follows that
\ber
<\vp_1,\vp_2> = \xi_1(v_2)+ \xi_2(v_1)
\eer
 defines a metric of signature $(d,d)$.
  
We shall further assume the existence of a generalised metric which, in matrix notation,  can be arranged to read
\ber
{\cal G}=\left( \begin{array}{cc}0&g^{-1}\\
g&0\end{array}\right)
\eer
corresponding to a metric splitting\footnote{ See \cite{Cavalcanti:2012fr} for the conditions for this to be possible in general.} of $\mathbb {T}$.
Introducing the projection operators 
\ber
\pi^g_\pm :=\half\left(1\pm {\cal G}\right)
\eer
the splitting reads
\ber
\mathbb {T}=\mathbb {T}_+\oplus\mathbb {T}_-~,
\eer
with
\ber
\mathbb {T}_\pm:=\pi^g_\pm\mathbb {T}~
\eer
being the $\pm 1$ eigenspaces of $\mathbb {T}$. The description of the SKT geometry will focus on the $+1$ eigenspace  $\mathbb {T}_+$.

\section{A (half) generalised complex structure on $\mathbb {T}_+$ }

A generalised almost complex structure ${\cal J}$ is an endomorphism of $\mathbb {T}$ which squares to minus the identity and preserves the natural pairing ${\cal P}$:
\ber\nn\label{jfull}
&&{\cal J}^2=-1\\[1mm]\nn
&&{\cal J}^t{\cal P}=-{\cal P}{\cal J}~.
\eer
Here we consider instead a half generalised structure;
a map ${\cal J}_+$ which acts on $\mathbb {T}_+$,  vanishes on $\mathbb {T}_-$, satisfying 
\ber\nn\label{cnds}
&&{\cal J}^2_+=-\pi^g_+\\[1mm]\nn
&&[{\cal J}_+,{\cal G}]=0\\[1mm]
&&(\pi^g_+)^t{\cal J}^t_+{\cal P}{\cal J}_+\pi^g_+=(\pi^g_+)^t{\cal P}\pi^g_+~.
\eer
Since ${\cal J}_+\pi^g_-=0$, we write it as
\ber
{\cal J}_+={\cal J}\pi^g_+~,
\eer
where  ${\cal J}$ is an almost complex structure on $\mathbb {T}$  as in \re{jfull}.
The second condition in \re{cnds} then implies the generalised Hermiticity condition
\ber\label{JG}
[{\cal J},{\cal G}]=0~,
\eer
and the first and third conditions in \re{cnds} then also hold.

\section{A coordinate description}

To describe the geometry, it is sometimes convenient to choose a coordinate basis $(\partial/\partial X^\mu, d X^\mu)$ for $\mathbb {T}$ and to write the elements of sections as
\ber\label{element}
\vp=\left(\begin{array}{c}v^\mu\\ \xi_\mu\end{array}
\right)
\eer
We then have a matrix description of our operators as
\ber
{\cal J}=\left( \begin{array}{cc}I&P\\
L&K\end{array}\right)
~,~~~\pi^g_+=\half\left( \begin{array}{cc}1&g^{-1}\\
g&1\end{array}\right)
~,~~~{\cal{P}}=\half\left( \begin{array}{cc}0&1\\
1&0\end{array}\right)
\eer
where
\ber\nn
&I=(I^\mu_{~\nu}): ~T{\cal M} \to T{\cal M} ~,~~~&P=(P^{\mu\nu}): ~T{\cal M} \to T^*{\cal M}\\[1mm]\nn
&L=(L_{\mu\nu}): ~T^*{\cal M} \to T{\cal M} ~,~~~&K=(K_\mu^{~\nu}): ~T^*{\cal M} \to T^*{\cal M}\\[1mm]
&g^{-1}=(g^{\mu\nu}):~T{\cal M} \to T^*{\cal M}~,~~~&g=(g_{\mu\nu}): ~T^*{\cal M} \to T{\cal M}~.
\eer
Using these, we find from \re{jfull} that ${\cal J}^2=-1$ implies
\ber\nn\label{cnd1}
&&I^2+PL=-1\\[1mm]\nn
&&IP+PK=0\\[1mm]\nn
&&LI+KL=0\\[1mm]
&&LP+K^2=-1~,
\eer
and that ${\cal J}^t{\cal P}=-{\cal P}{\cal J}$ implies
\ber\nn\label{cnd2}
&&P^t+P=0\\[1mm]\nn
&&L^t+L=0\\[1mm]
&&I+K^t=0~.
\eer
Finally, the condition \re{JG} that $[{\cal J},{\cal G}]=0$ gives
\ber\nn\label{cnd3}
&&Pg-g^{-1}L=0\\[1mm]
&&Kg-gI=0~.
\eer
The conditions \re{cnd1}-\re{cnd3} may be summarised by saying that ${\cal J}$ may be written
\ber
{\cal J}=\left( \begin{array}{cc}\hat I\mp Pg&P\\
gPg&-\hat I^t \pm gP\end{array}\right)
\eer
where $P$ is antisymmetric and $\hat I$ is an almost complex structure on ${\cal M}$ that preserves the metric:
\ber\label{Icon}
P=-P^t~,~~~\hat I:=I\pm Pg~,~~~(\hat I)^2=-1~,~~~(\hat I)^tg\hat I=g~.
\eer
It follows that the complex structure  ${\cal J}_+$ on $\mathbb {T}_+$ is
\ber\label{finalJ}
{\cal J}_+={\cal J}\pi^g_+=\pi^g_+{\cal J}\pi^g_+=\half\left( \begin{array}{cc}\hat I&-(\hat\omega)^{-1}\\
\hat\omega&-\hat I^t \end{array}\right)
\eer
where $\hat \omega:=g\hat I$.

\section{Integrability}
\label{inte}

Using ${\cal J}_+$ from \re{finalJ}, we define another two projection operators
\ber\label{jproj}
\pi^j_\pm:=\half (1\pm i{\cal J}_+)~.
\eer
This allows a further split
\ber
\mathbb {T}_+ = {\mathbb {T}}^{(1,0)}_+\oplus {\mathbb {T}}^{(0,1)}_+
\eer
where
\ber
 {\mathbb {T}}^{(1,0)}_+=\pi^j_+  {\mathbb {T}}_+~,~~~ {\mathbb {T}}^{(0,1)}_+=\pi^j_-  {\mathbb {T}}_+
\eer
are  $+i$ and $-i$ eigenspaces, respectively. We then require ${\mathbb {T}}^{(1,0)}_+$  to be involutive with respect to the $H-$twisted Courant bracket, which reads
\ber\label{courant}\nn
&&\llbracket \vp_1,\vp_2\rrbracket_H\\[1mm]
&&:=\left(\begin{array}{c}[v_1,v_2]\\ \Lie_{v_1}d\xi_2 -i_{v_2}d\xi_1+i_{v_1}i_{v_2}H\end{array}
\right)=\left(\begin{array}{c}[v_1,v_2]\\ 
2i_{v_{[1}}d\xi_{{}_{2]}}
+d(i_{v_1}\xi_2)
+i_{v_1}i_{v_2}H\end{array}
\right)~,
\eer
with $dH=0$.
When $\vp_i\in {\mathbb {T}}^{(1,0)}_+$, they may be written
\ber\label{rightT}
\vp_i=\pi^j_+\pi^g_ +\vp_i~,~\iff ~\vp_i=\half\left(\begin{array}{c}p_+\hat v_i\\\ gp_+\hat v_i\end{array}\right)~.
\eer
where
\ber\nn
&&p_+:=\half(1+i\hat I)\\[1mm]
&&\hat v_i:=v_i+g^{-1}\xi_i
\eer
The involution conditions translate into
\ber\label{courg1}
\pi^g_-\llbracket \vp_1 ,\vp_2\rrbracket_H=0~, \iff ~~<\tilde \vp_3, \llbracket \vp_1 ,\vp_2\rrbracket_H>=0~,~~\tilde \vp_3\in {\mathbb {T}}_-
\eer
and 
\ber\label{courj}
\pi^j_-\llbracket \vp_1 ,\vp_2\rrbracket_H=0 ~\iff~~<\tilde \vp_3 ,\llbracket \vp_1 ,\vp_2\rrbracket_H>=0~,~~\tilde \vp_3\in {\mathbb {T}}^{(0,1)}_+
\eer
\noindent
to stay in ${\mathbb {T}}^{(1,0)}_+$. (Note that $[\pi^g,\pi^j]=0$). The first condition leads to ${\cal {J}}_+$ being parallel, the second to the vanishing of the Nijenhuis tensor for $I$.

Letting $p_+ \hat v_i=:w_i$, we learn from \re{courg1}  that 
\ber\label{zucchini}
[w_1,w_2]^\mu-g^{\mu\nu}\left(2i_{w_{[1}}d(gw)_{{}_{2]}}
+d(i_{w_1}gw_2)
+i_{w_1}i_{w_2}H\right){}_\nu=0~.
\eer
For $\vp_i\in  {\mathbb {T}}_+ : \xi_i=gv_i$ we have 
\ber
\left(2i_{v_{[1}}d(gv)_{{}_{2]}}
+d(i_{v_1}gv_2)
+i_{v_1}i_{v_2}H\right)_\mu=g_{\mu\kappa}[v_1,v_2]^\kappa+2v_{2\nu}\nabla_\mu^{(+)}v_1^\nu
\eer
where $\nabla_\mu^{(+)}$ is the covariant derivative with torsion
\ber\label{cov+}
\nabla_\mu^{(+)}v_1^\nu=\nabla_\mu^{(0)}v_1^\nu +\half H_{\mu\rho}^{~~\nu}v^\rho_1~,
\eer
with $\nabla_\mu^{(0)}$ the Levi-Civita connection for $g$.
Using this in \re{courg1} yields
\ber\label{wdw}
w_{2\nu}\nabla_\mu^{(+)}w_1^\nu=0~.
\eer
Since $p_+gp_+=0$, we can peel off the $\hat v_i$\! s to conclude that
\ber\label{abel}
\nabla^{(+)}_\mu \hat I^\tau_{~\nu}=0~.
\eer
On $T{\cal M}$ the complex structure $\hat I$ is thus parallel with respect to this torsionful connection.
From \re{courj} we find
\ber
&&(\half+p_-)[w_1,w_2]+\ihalf (\hat \omega)^{-1}\left(2i_{w_{[1}}d(gw)_{{}_{2]}}
+d(i_{w_1}gw_2)
+i_{w_1}i_{w_2}H\right)=0~,\\[2mm]
&&-\ihalf\hat\omega [w_1,w_2]+(\half+p_+^t)\left(2i_{w_{[1}}d(gw)_{{}_{2]}}
+d(i_{w_1}gw_2)
+i_{w_1}i_{w_2}H\right)=0~.
\eer
From these and \re{zucchini} it follows that
\ber\nn
&&p_-[w_1,w_2]= p_-[p_+\hat v_1,p_+\hat v_2]=0~,\\[1mm]
&&p_-g^{-1}\left(2i_{w_{[1}}d(gw)_{{}_{2]}}
+d(i_{w_1}gw_2)
+i_{w_1}i_{w_2}H\right)\\[1mm]
&&~~~~~~~~~~~~~=p_-[w_1,w_2]+2p_-i_{gw_{2}}g^{-1}\nabla^{(+)}w_1=0~.
\eer
The first of these is the integrability condition for $\hat I$ on ${\cal M}$:
\ber\label{nij}
{\cal N}(\hat I)=0~,
\eer
where ${\cal N}$ is the Nijenhuis tensor. The second relation follows from the first and \re{wdw}.

From the vanishing of the Nijenhuistensor, \re{nij} in conjunction with the parallel condition  \re{abel},  one derives the torsion relation
\ber
T_{\mu\nu\rho}=T_{\sigma\tau [\rho}\hat I^\sigma_{~\mu}\hat I^\tau_{~\nu]}~,
\eer
which is equivalent to the final condition
\ber
d^c\hat\omega = H~,
\eer
for $T=\half H$.  Note tha $H$ is closed by assumption which means that
\ber
dd^c\hat\omega = 0~.
\eer

\section{Generalised K\"ahler Geometry}

Generalised K\"ahler Geometry (GKG) is the target space geometry of $(2,2)$ sigma models. It is determined by two complex structures $\hat I^{(\pm)}$, a metric $g$ which is hermitian with respect to both of these and a closed three form $H$ which enters the conditions for integrability. The relation to GCG is given by the Gualtieri map \cite{Gualtieri:2003dx}
\ber\label{Gmap}
{\cal{J}}^{(1,2)}=
\half\left(\begin{array}{cc}
\hat I^{(+)}\pm \hat I^{(-)}&-(\omega_{(+)}^{-1}\mp\omega_{(-)}^{-1})\cr
\omega_{(+)}\mp\omega_{(-)}&-(\hat I^{t(+)}\pm \hat I^{t(-)})\end{array}\right)
\eer{}
where ${\cal{J}}^{(1)}$ and ${\cal{J}}^{(2)}$ are two commuting Generalised Complex Structures with integrability defined with respect to the $H$-twisted Courant bracket \re{courant}. GKG  has been extensively studied in the context of sigma models \cite{Buscher:1987uw}-\cite{Bischoff:2018kzk}, but here we just want to elucidate the relation to the half generalised complex structures discussed above.

Half generalised complex structures can be defined on  $\mathbb {T}_- $ in the same way as described for $\mathbb {T}_+ $. Assume that 
 ${\cal J}_+$ is defined on  $\mathbb {T}_+$ and takes the form \re{finalJ}
 \ber
{\cal J}_+={\cal J}\pi^g_+=\pi^g_+{\cal J}\pi^g_+=\half\left( \begin{array}{cc}\hat I^{(+)}&-(\hat\omega_{(+)})^{-1}\\
\hat\omega_{(+)}&-\hat I^{(+)t}\end{array}\right)~,
\eer
and  ${\cal J}_-$ is defined on  $\mathbb {T}_-$ 
 \ber
{\cal J}_-=\tilde{\cal J}\pi^g_-=\pi^g_-\tilde{\cal J}\pi^g_-=\half\left( \begin{array}{cc}\hat I^{(-)}&-(\hat\omega_{(-)})^{-1}\\
\hat\omega_{(-)}&-\hat I^{(-)t}\end{array}\right)~,
\eer
both integrable with respect to the same $H$-twisted Courant bracket. Their sum and difference then yield precisely \re{Gmap}
\ber
{\cal{J}}^{(1,2)}=\pi^g_+{\cal J}\pi^g_+\pm \pi^g_-\tilde{\cal J}\pi^g_-~.
\eer
At the sigma model level this is mirrored by the fact that a  $(2,2)$ model can be thought of as the sum of a  $(2,0)$ and a  $(0,2)$ model.

\section{Conclusions}

We have briefly described how SKT geometry fits into Generalised Complex Geometry as half generalised structures, related it to $(2,0)$ and $(2,1)$ sigma model target space geometry and shown how two half generalised complex structures can give rise to the Generalised K\"ahler Geometry of $(2,2)$ sigma models. 
These considerations generalise to $(p,q)$ supersymmetric models and their target space geometries as shown in  \cite{Hull:2018jkr}.

\vspace{2cm}

\noindent{\bf Acknowlegement}: \\
We have benefitted from comments on the manuscript by Gil Cavalcanti.
UL gratefully acknowledges the hospitality of the theory group at Imperial College, London. 
 This work was supported by the EPSRC programme grant ``New
Geometric Structures from String Theory'',  EP/K034456/1.

\end{document}